\documentclass[preprint,12pt]{elsarticle}

\usepackage{graphics}

\usepackage{amssymb}

\journal{Journal of Crystal Growth}

\begin{document}

\begin{frontmatter}



\title{Monte Carlo simulation of growth of hard-sphere crystals on a square pattern}


\author{Atsushi Mori}

\address{Institute of Technology and Science, The University of Tokushima,
Tokushima 770-8506, Japan}

\begin{abstract}
Monte Carlo simulations of the colloidal epitaxy of hard spheres (HSs) on a square pattern have been performed.
This is an extension of previous simulations; we observed a shrinking intrinsic stacking fault running in an oblique direction through the glide of a Shockley partial dislocation terminating its lower end in fcc (001) stacking [Mori et~al., Molec. Phys. {\bf 105} (2007) 1377], which was an answer to a question why the defect in colloidal crystals reduced by gravity [Zhu et~al., Nature {\bf 387} (1997) 883].
We have resolved one of shortcomings of the previous simulations; the driving force for fcc (001) stacking, which was stress from a small periodic boundary simulation box, has been replaced with the stress from a pattern on the bottom.
We have observed disappearance of stacking fault in this realizable condition.
Sinking of the center of gravity has been smooth and of a single relaxation mode under the condition that the gravitational energy $mg\sigma$ is slightly less than the thermal energy $k_{\mbox{\scriptsize B}}T$.
In the snapshots tetrahedral structures have appeared often, suggesting formation of staking fault tetrahedra.
\end{abstract}


\end{frontmatter}


\section{Introduction}
\label{sec:intro}
In 1957 the crystalline phase transition was discovered in the hard-sphere (HS) system by a Monte Carlo (MC) simulation \cite{Woodcock1957} and a molecular dynamics (MD) simulation \cite{Alder1957}.
Their results were surprising because the phase transition occurred in a pure repulsive system.
In 1960-70s colloidal crystallizations were extensively studied as the HS crystalline phase transition in reality.
For historical details, see, for example, the introduction of a review \cite{MoriTBP}.
Recent situation of studies on the colloidal crystal is different from that in those days; so-called HS suspensions are synthesized \cite{Antl1986}, which exhibit a HS nature in the crystal-fluid phase transition \cite{Pusey1986,Paulin1990,Underwood1994,Phan1996}.
There is another trend of studies of the colloidal crystals in recent days.
Because in colloidal crystals a periodic structure of dielectric constant with the periodicity of the same order of optical wavelength, the colloidal crystals can be used as photonic crystals \cite{Ohtaka1979,Yablonovitch1987,John1987}.
As compared to micro manufacturing technologies of fabricating the photonic crystals, the colloidal crystallization is of low cost in introducing equipment and less time consuming in the fabrication.
One of shortcomings of the colloidal crystallization is that the colloidal crystals contain many crystal defects.
From fundamental as well as application point of view, the defect in the photonic crystal should be reduced.
The photonic band cannot be opened unless the defect is reduced.

In relation to the reduction of the crystal defect in the colloidal crystals, in 1997 Zhu et~al. \cite{Zhu1997} found an effect of gravity that reduces the stacking disorder in the HS colloidal crystals.
They found that the colloidal particles formed a random hexagonal close pack (rhcp) structure under micro gravity.
On the other hand, the sediment is rhcp/face-center cubic (fcc) mixture under normal gravity \cite{Pusey1989}.
The mechanism of the reduction of the stacking disorder under gravity was so far unresolved until the present author and coworkers found a glide mechanism of the disappearance of a stacking fault \cite{Mori2007MP}.
Viewing $\langle$111$\rangle$ fcc is characterized by a stacking of the ABCABC$\cdots$ sequence, where A, B, and C distinguish hexagonal planes on the basis of the positions of the particles in the hexagonal plane.
On the other hand, hexagonal close pack (hcp) structure is given by ABAB$\cdots$ stacking and rhcp by a random sequence of A, B, and C.
The stacking disorder is the disorder in the sequence of A, B, and C.
For example, an intrinsic stacking fault is given by a sequence such as ABABC$\cdots$; here the third C plane has been removed from ABCABC$\cdots$.
We note that even if the stacking is out of order, the particle number density remains unchanged.
In this respect, the varieties of stacking sequence are not affected by gravity.
So, the mechanism of the reduction of the stacking disorder due to gravity was a long standing problem.

In Ref.~\cite{Mori2007MP} looking into the evolution of snapshots of MC simulations of HSs \cite{Mori2006JCP}, in which transformation from a defective crystal into a less-defective crystal under gravity was observed, we found that a glide of a Shockley partial dislocation terminating an intrinsic stacking fault shrunk the stacking fault in fcc (001) stacking.
The key is the fcc (001) stacking; in those simulations this stacking was forced due to a stress from a small periodic boundary simulation box.
In contrast, in the colloidal crystallization patterned bottom walls are sometimes used; the fcc (001) stacking is forced due to the stress from the pattern on the bottom.
Use of the patterned bottom wall is called a colloidal epitaxy.
In 1997 van Blaaderen et~al. succeeded in the fcc (001) stacking using a fcc (001) pattern \cite{Blaaderen1997}.
The basic idea of the colloidal epitaxy is that the stacking sequence is unique in $\langle$100$\rangle$.
The finding of Ref.~\cite{Mori2007MP} is that in the fcc (001) stacking, even if an intrinsic stacking fault running along oblique \{111\} plane is introduced, through the glide of a Shockley partial dislocation terminating the lower end of the stacking fault the stacking fault shrinks.
In other words, Ref.~\cite{Mori2007MP} points out a superiority of the colloidal epitaxy other than the unique stacking sequence.
We note here that this glide mechanism is merely one of mechanisms.
The intrinsic stacking fault is mere one of metastable configurations.
Therefore, there exist mechanisms connecting other metastable configurations.
Moreover, we have already found a configuration which was succeeded in newly grown crystal in the fluid phase in simulations of the same condition \cite{Mori2007FPE}.
In addition, we confirmed that a coherent growth occurred in the simulations \cite{Mori2006STAM}.
Complementarily to the simulations, we have given elastic energy calculations to understand the driving force of upward move of the Shockley partial dislocation \cite{Mori2009,Mori2010}.

The purpose of the present simulation is to resolve the shortcoming of previous simulations \cite{Mori2007MP,Mori2006JCP,Mori2007FPE,Mori2006STAM}.
In those simulation fcc (001) stacking was forced due to the stress from a small periodic boundary simulation box.
This artifact should be resolved.
Of course, the same stress can be provided by the patterned substrate (the colloidal epitaxy).
However, the system size cannot be systematically enlarged in the previous simulations.
As already shown \cite{Mori2006JCP} fcc \{111\} stacking occurs for a large lateral system size.
In the present simulation we use a square pattern.
An advantage of the square pattern is that matching between the crystal grown and the substrate on the lattice line, not only on the lattice point, is possible \cite{Lin2000}.

\section{Simulation method}
\label{sec:method}
HSs (diameter $\sigma$) under gravity (the acceleration due to gravity $g$) were confined in a simulation box with the periodic boundary condition in horizontal direction and a top flat and bottom square-patterned hard walls (Fig.~1 of Ref.~\cite{MoriTBP}).
The groove width was $0.707106781\sigma$.
So, the diagonal distance of the intersection of the longitudinal and transverse grooves was $0.707106781\sigma \times \sqrt{2} = 0.9999999997\sigma$.
Thus, a HS located on the lattice point of the bottom square lattice fell into the intersection of the grooves by almost the half of HS diameter.
The separation between neighboring groove edges was $0.338\sigma$.
Accordingly, the periodicity of the lattice was $1.045\sigma$ and the diagonal distance $1.478\sigma$; i.e., we set the bottom lattice so as to coincide with the bottom (001) layer of the fcc crystal of the previous flat wall simulation \cite{Mori2006STAM}.


In this paper we shall take some close looks into the simulation results of two lateral system sizes, $L_x$ = $L_y$ = $12.55\sigma$ and $L_x$ = $L_y$ = $25.09\sigma$.
The vertical system size was fixed at $L_z$ = $200\sigma$; this size was enough large so that at the initial ($g^*$ = 0.0) the HSs were dispersed randomly.
To prepare an initial state we ran a MC simulation for 2$\times 10^7$ MC cycle (MCC).
Here, one MCC was define so that it contains $N$ MC particle moves, i.e., every particle undergoes one particle move on average in one MCC.
The maximum displacement was fixed at $\Delta r_{\mbox{\scriptsize max}}$ = $0.06\sigma$ throughout.
Therefore, as the density changes the acceptance ratio changes; as a result the time corresponding to one MCC varied as the simulation proceeded.
Nevertheless, the states emerges in a course of a simulation are arranged as a time series.
The numbers of particles were $N$ = 6656 and 26624.
These numbers of particle, $N$s, were selected such that the numbers of particles laid on the bottom per unit area, $n_{\mbox{\scriptsize s}}$, became the same value as that of previous simulations \cite{Mori2006JCP}.


In those simulations the gravitational number $g^* \equiv mg\sigma/k_{\mbox{\scriptsize B}}T$ was increased stepwise to avoid the trapping of the system by a metastable state such as polycrystalline state \cite{Mori2006JCP}.
Here, $m$ is the mass of a particle, $k_{\mbox{\scriptsize B}}T$ the temperature multiplied by Boltzmann's constant.
If gravity such as $g^*$ = 0.9 is suddenly switched on for the flat wall case, the system is trapped into a polycrystalline metastable state \cite{Yanagiya2005}.
We note here that an effective control of $g^*$ can be done in a centrifugation method \cite{Suzuki2007}, in comparison to a gravitational sedimentation.
Parameters of the stepwise control of $g^*$ was $\Delta t$ = $2\times10^5$MCC for $N$ = 6656 system and $\Delta t$ = $8\times10^5$MCC for $N$ = 26624; $\Delta g^*$ = 0.1 for both.
In Ref.~\cite{Mori2006JCP} at first we kept $g^*$ at 0 for $\Delta t$ and then increased $g^*$ by $\Delta g^*$.
On the other hand, in this simulation we kept $g^*$ at $\Delta g^*$ at first for $\Delta g$ and then increased.
Although several combinations were tested, we will not present on the optimization of $g^*$ control in this paper for the limitation of pages.
Without any optimization we have got results, which were enough to complement the shortcoming of the previous studies.
With an optimization we will look at detailed processes of the defect disappearance.
With other optimizations disappearance of several types of defects must be observed.

\section{Results and discussions}
\label{sec:result}
We have preformed seven simulations for $N$ = 6656 system and three for $N$ = 26624 with different series of random numbers.
In two of these simulations for $N$ = 6656 defect disappearance at $g^*$ less than 0.9 was observed; at $g^*$=0.9 shrinking of an intrinsic stacking fault occurred for a flat bottom wall case \cite{Mori2007MP}.
In four of these simulations for $N$ = 6656 defect disappearance occurred at $g^*$ greater than 0.9.
For remainder one defect disappearance was not appreciable.
For $N$ = 26624 system in all three simulation defect disappearance was observed at $g^*$ less than 0.9.
In two of three the defect disappearance occurred during $g^*$ = 0.5 and in the remainder one during $g^*$ = 0.7.
What was notable for $N$ = 26624 system was that increase of disorder in appearance was seen at large $g^*$.
In a case this phenomenon was observed in one of projected snapshot and, on the other hand, defect disappearance was observed in the other projection.
Let us postpone the detail analysis and discussion after looking at snapshots.
Snapshots and evolutions of the center of gravity will be shown in section \ref{sec:6656} for $N$ = 6656 system and in section \ref{sec:26624} for $N$ = 26624.

\subsection{$N$=6656 system}
\label{sec:6656}
\begin{figure}[htb]
\begin{center}
\includegraphics[width=12cm]{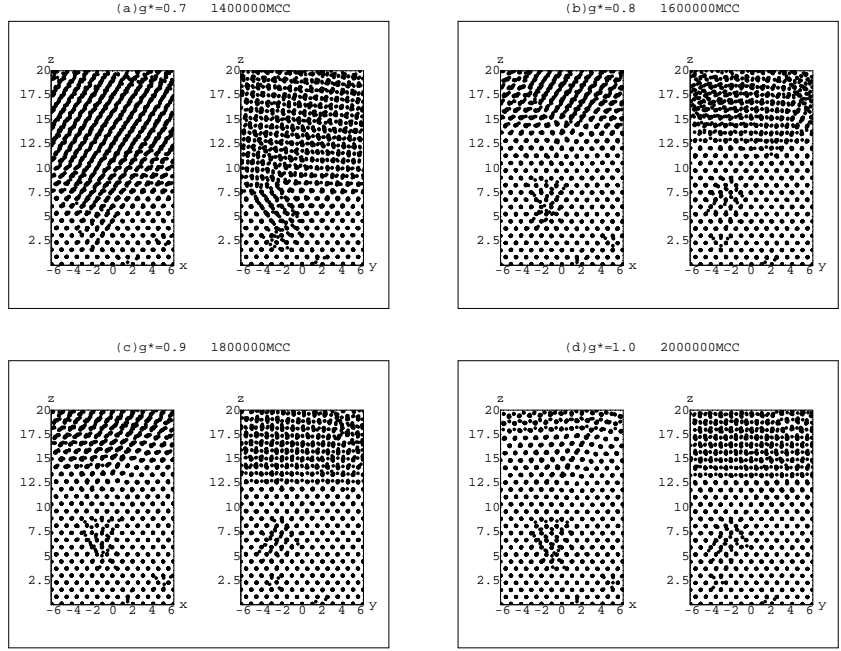}
\end{center}
\caption{Snapshots projected on $xz$ and $yx$ planes at (a) $g^*$=0.7, (b) 0.8, (c) 0.9, and (d) 1.0 for a case that the defect disappearance occurred at $g^*$ less than 0.9 for $N$ = 6656 system.
Defect existed in $7.5 < z/\sigma < 13.5$, as indicated by $yz$ projection of (a), disappeared during $g^*$ = 0.8.
Also defect in $13.5 < z/\sigma < 17.5$ in $xz$ projection of (c) disappeared during $g^*$ = 1.0.
Defects around $(x/\sigma,y/\sigma,z/\sigma)$ = $(-2,-3,6)$ and $(-5,-4,3)$ did not disappear.}
\label{fig:snaplow}
\end{figure}

Snapshots at $g^*$ = 0.7-1.0 are shown in Fig.~\ref{fig:snaplow} for a case that the defect disappearance occurred at $g^*$ less than 0.9.
Though defects in the lower portion remained, a defect in appearance expanded over the middle portion disappeared during $g^*$ = 0.8 and then that in top portion disappeared during $g^*$ = 1.0 again.
In fcc (001) stacking, if a single stacking fault runs along one of \{111\}, [110] ([1$\bar{1}$0]) lattice line makes an array of two separated points in (110) [(1$\bar{1}$0)] projection.
And, on the other projection we can observe a fault directly.
To understand Fig.~\ref{fig:snaplow} we must take into account the fact that $x$ and $y$ directions correspond to $\langle$110] (see Ref.~\cite{Mori2006JCP}).
In Fig.~\ref{fig:snaplow} (a) splittings are observed in both $xz$ and $yz$ projections and the both splittings disappeared in Fig.~\ref{fig:snaplow} (b).
Therefore, we cannot conjecture those as involving shrinking of a single stacking fault as for a flat wall case \cite{Mori2007MP}.
If two stacking faults along, {\it e.g.}, (111) and (11$\bar{1}$), coexist, then splitting on (110) projection and two intersecting fault on (1$\bar{1}$0) projection are seen.
So, two stacking faults along, {\it e.g.}, (111) and (1$\bar{1}$1), must coexist.
To observe intersections between (110) or (1$\bar{1}$0) and stacking faults by making three-dimensional (3D) view may give an answer.
The surface structure of the 3D snapshot was, however, complicated as imagined geometrically.
Although a crossing two faults was seen, we cannot successfully follow the evolution
as previously done \cite{Mori2007MP} because of the complexity.
Let us postpone this complex analysis as a future research.
On the other hand, comparing Fig.~\ref{fig:snaplow} (c) and (d) we find that splitting of $xz$ projection of lattice lines disappeared during $g^*$ = 1.0.
That the splitting in $yz$ projection did not disappear suggests that a single stacking fault such as running along, {\it e.g.}, (1$\bar{1}$1), remained.
In other words, disappearance of a stacking fault along (111) or (11$\bar{1}$) is deduced.

We discuss about the possibility of staking fault tetrahedra.
We observe downward triangles in $xz$ projection and upward triangles in $yz$ projection at the defect around $(x/\sigma,y/\sigma,z/\sigma)$ = $(-2,-3,6)$ in Fig.~\ref{fig:snaplow} (b)-(d).
Those triangles are seen if we make projections of a tetrahedron surrounded by \{111\} onto (110) and (1$\bar{1}$0).
Regarding the defect around $(x/\sigma,y/\sigma,z/\sigma)$ = $(-5,-4,3)$ we cannot identify the three dimensional shape.
Those defects seem to be sessile, because they remain for a long time.

\begin{figure}[htb]
\begin{center}
\includegraphics[width=5.0cm]{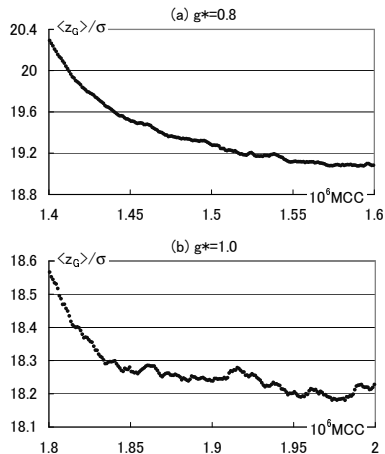}
\end{center}
\caption{Evolution of the center of gravity during (a) $g^*$=0.8 and (b) 1.0 for a case that the defect disappearance occurred at $g^*$ less than 0.9 for $N$ = 6656 system.
The curve in (a) is more smooth as compared to that in Ref.~\cite{Mori2007MP}.
That in (b), however, exhibits a multiple relaxation manner as in Ref.~\cite{Mori2007MP}.
Statistical errors are within $0.011\sigma$ for (a) and $0.008\sigma$ for (b).}
\label{fig:gcenterlow}
\end{figure}

Figure~\ref{fig:gcenterlow} is the evolution of the center of gravity for a case that the defect disappearance occurred at $g^*$ less than 0.9.
For the flat wall case we observed some plateaus in the evolution of the center of gravity.
In contrast, the evolution of the center of gravity in the preset case during $g^*$ = 0.8 is more smooth and nearly of a single relaxation mode.
On the other hand, that during $g^*$ = 1.0 implies trapping at a metastable configuration during defect disappearance.
The relaxation in Fig.~\ref{fig:gcenterlow} (b) is of two step manner.
In 1.87-1.9 $\times10^6$MCC a settlement at a metastable configuration occurred and then a relaxation started again.

\begin{figure}[htbp]
\begin{center}
\includegraphics[width=12cm]{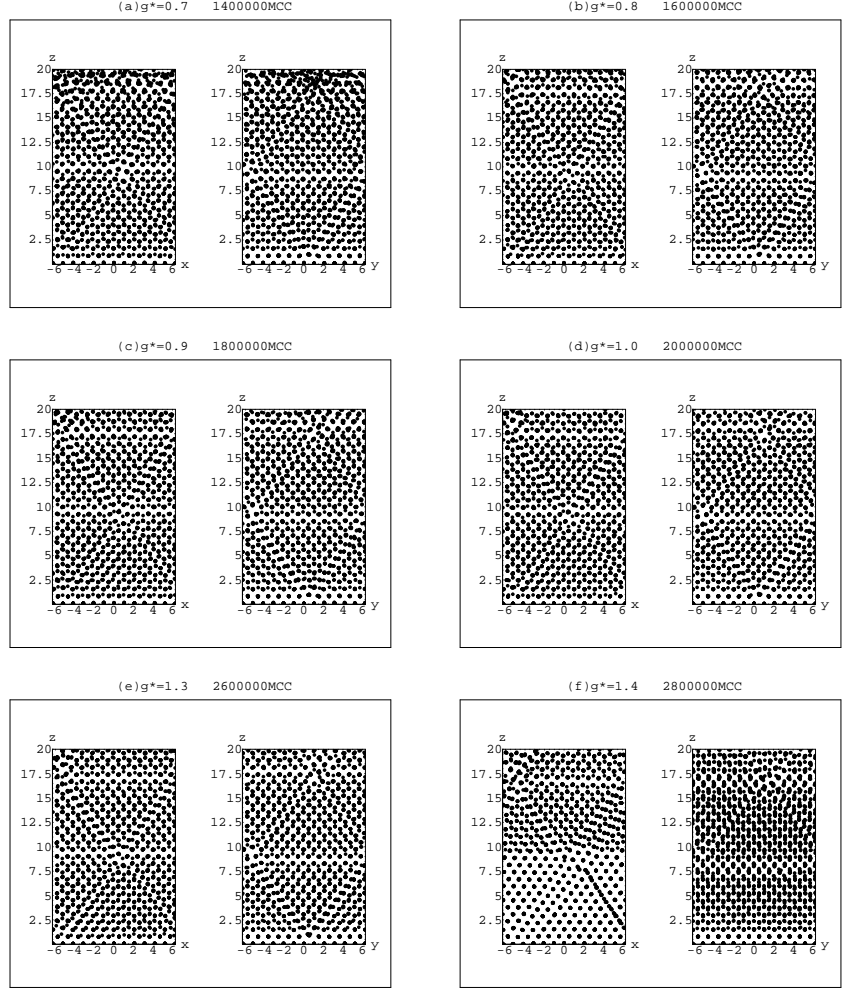}
\end{center}
\caption{Snapshots projected on $xz$ and $yx$ planes at (a) $g^*$=0.7, (b) 0.8, (c) 0.9, (d) 1.0, (e) 1.3, and (f) 1.4 for a case that the defect disappearance occurred at $g^*$ greater than 0.9 for $N$ = 6656 system.
Whereas no defect disappearance occurred in (a)-(d), defect in $2<z/\sigma<9$ in $xz$ projection of (e) disappeared in (f) during $g^*$=1.4.}
\label{fig:snaphigh}
\end{figure}

Let us look at snapshots for a case that the defect disappearance occurred at $g^*$ greater than 0.9.
Snapshots are shown in Fig.~\ref{fig:snaphigh}.
At first, we note that two layer defect free region are formed at the bottom.
We observed formation of a few crystalline layers commonly for seven simulations performed at low $g^*$ such as 0.5.
Looking at snapshots (not shown) of those layers parallel to the bottom wall, we find that the bottom layers match to the pattern on the bottom wall on the lattice point.
From density profiles (not shown) we find that layering of two layers, which possessed no significant interlayer ordering, at the bottom occurred even at $g^*$ = 0.2.
This layering phenomena are common to flat wall cases \cite{Biben1994,Marechal2007}.
Defect disappearance did not occurred up to $g^*$ = 1.4.
During $g^*$ = 1.4 defect shown in $xz$ projection disappeared.
We confirm a stacking fault in the right-lower region in $xz$ projection in Fig.~\ref{fig:snaphigh} (e).
In addition, following a lattice line horizontally we find a step on a lattice line in the left-lower region.
This is characteristic of a stacking fault.
Accordingly, there exist two stacking faults of different directions.
At ($x/\sigma$,$z/\sigma$) $\sim$ (2,7.5) those two staking faults meet.
There is a possibility of a star-rod partial dislocation there.

Let us look at the portion $9 < z/\sigma < 17$, {\it e.g.}, in Fig.~\ref{fig:snaphigh} (f).
We observe a downward triangle in $xz$ projection and an upward triangle in $yx$ projection at the middle.
Taking into account the periodic boundary condition, both sides of these triangles make triangles upward and downward.
It is suggested that an upward tetrahedron and downward tetrahedron fills a part of the space.
However, this situation takes place so as to match the periodic boundary condition.
So, if the system size is large enough, a configuration such that a tetrahedron is embedded in a defect free matrix as Fig.~\ref{fig:snaplow} (b)-(d) must be seen.
Why tetrahedral configuration did not appeared in the flat wall cases might be due to the system size.

\begin{figure}[htb]
\begin{center}
\includegraphics[width=5.0cm]{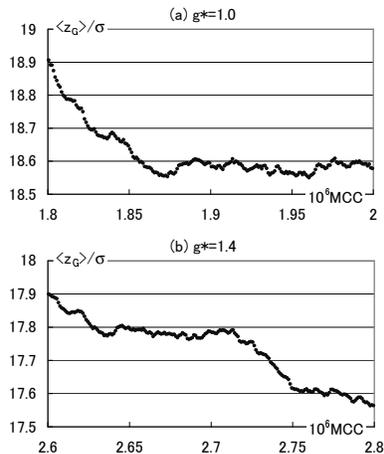}
\end{center}
\caption{Evolution of the center of gravity during (a) $g^*$=1.0 and (b) 1.4 for a case that the defect disappearance occurred at $g^*$ greater than 0.9 for $N$ = 6656 system.
The curve in (a) is of a single relaxation.
On the other hand, that in (b) exhibits a multiple relaxation manner as in Ref.~\cite{Mori2007MP}.
Statistical errors are within $0.008\sigma$ for (a) and $0.005\sigma$ for (b).}
\label{fig:gcenterhigh}
\end{figure}

We have compared the evolution of the center of gravity for this case and that for a case that the defect disappearance occurred at $g^*$ less than 0.9 during $g^*$ = 0.8.
There is no significant difference, though, if the defect disappearance did not occur due to trapping at the metastable configuration, plateaus indicating this trap were expected.
We speculate that upward growth of nucleated crystalline layers on the bottom, which occurred up to $g^* \sim 0.4$ as mentioned above, is involved in this smooth relaxation.
Figure~\ref{fig:gcenterhigh} is the evolution of the center of gravity for a case that the defect disappearance occurred at $g^*$ greater than 0.9.
It is interesting that both in Fig.~\ref{fig:snaplow} (d) and Fig.~\ref{fig:snaphigh} (f) the relaxation is multiple manner when the process is from a state including ^^ ^^ multiple" stacking faults to that including a ^^ ^^ single" stacking fault.

\subsection{$N$=26624 system}
\label{sec:26624}

\begin{figure}[htb]
\begin{center}
\includegraphics[height=15cm]{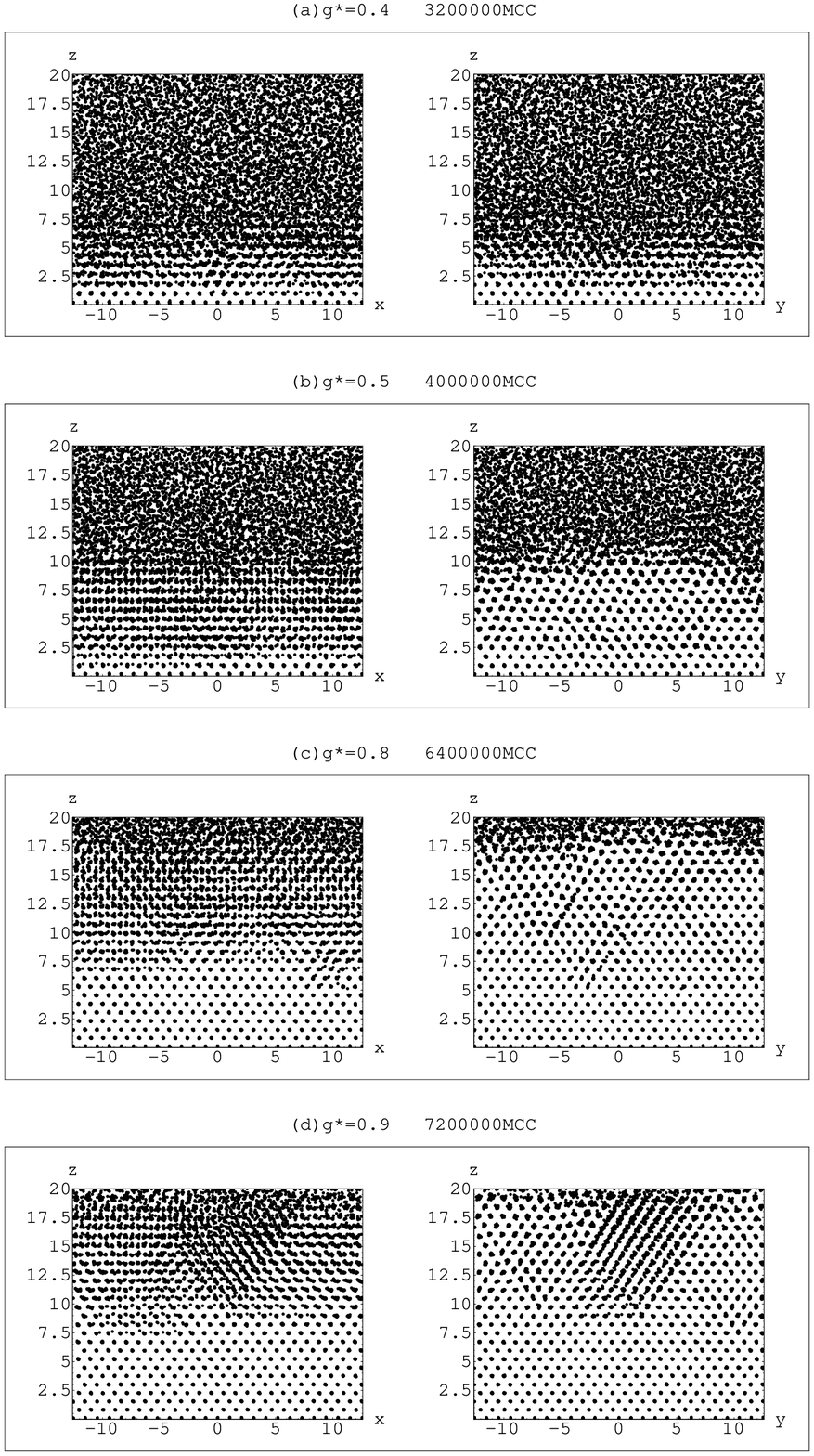}
\end{center}
\caption{Snapshots projected on $xz$ and $yx$ planes at (a) $g^*$=0.4, (b) 0.5, (c) 0.8, and (d) 0.9 for $N$ = 26624 system.
Defect disappearance occurred in $xz$ projection of (a)-(b) and (c)-(d), and increasing of disorder in appearance is observed in $yz$ projection of (c)-(d).}
\label{fig:snap26624}
\end{figure}

Snapshots at $g^*$ = 0.4-0.5 and 0.8-0.9 are shown in Fig.~\ref{fig:snap26624}.
Looking at the $yz$ projection of Fig.~\ref{fig:snap26624} (a) and (b) we see that defect in $3 < z/\sigma < 9$ disappeared during $g^*$ = 0.5.
Splitting of the projection of lattice lines in $xz$ projection of Fig.~\ref{fig:snap26624} (b) in this region implies that lattice lines along $y$ axis are crossing with faults.
Indeed, we can confirm steps on a lattice line traversing along the lattice line horizontally in $yx$ projection of Fig.~\ref{fig:snap26624} (b).
As for the $N$ = 6656 system disappearance of stacking faults in one direction is suggested.
Let us compare $xz$ projections of Fig.~\ref{fig:snap26624} (b) and (c).
We find that some splittings of projections of lattice lines disappeared, indicating that disappearance of the stacking fault in the corresponding direction.
We note that new defects such as at $(y/\sigma,z/\sigma)$ = $(-6,10)$-$(-4,13)$, around $(y/\sigma,z/\sigma)$ = $(-3,5)$-$(-1,8)$ formed.
The former may be a stacking fault.
Indeed splitting of projections of lattice lines is seen in $xz$ projection of Fig.~\ref{fig:snap26624} (c) over the levels same at this defect in $yz$ projection.
On the other hand, the latter defect is somewhat widened.
We cannot identify only from the projected snapshots.

Let us look at Fig.~\ref{fig:snap26624} (d).
A defect is expanded over a wide region in $yz$ projection.
This is a newly formed defect.
Correspondingly, we see a downward thick triangle structure and an upward thick triangle structure in $xz$ projection.
If a downward (upward) triangle in one direction corresponds to an upward (downward) triangle in the other projection, a stacking fault tetrahedron is suggested.
Simultaneously, we observe defect disappearance around $(x/\sigma,z\sigma)$ = $(11,6)$ in $xz$ projection.

\begin{figure}[htb]
\begin{center}
\includegraphics[width=5.0cm]{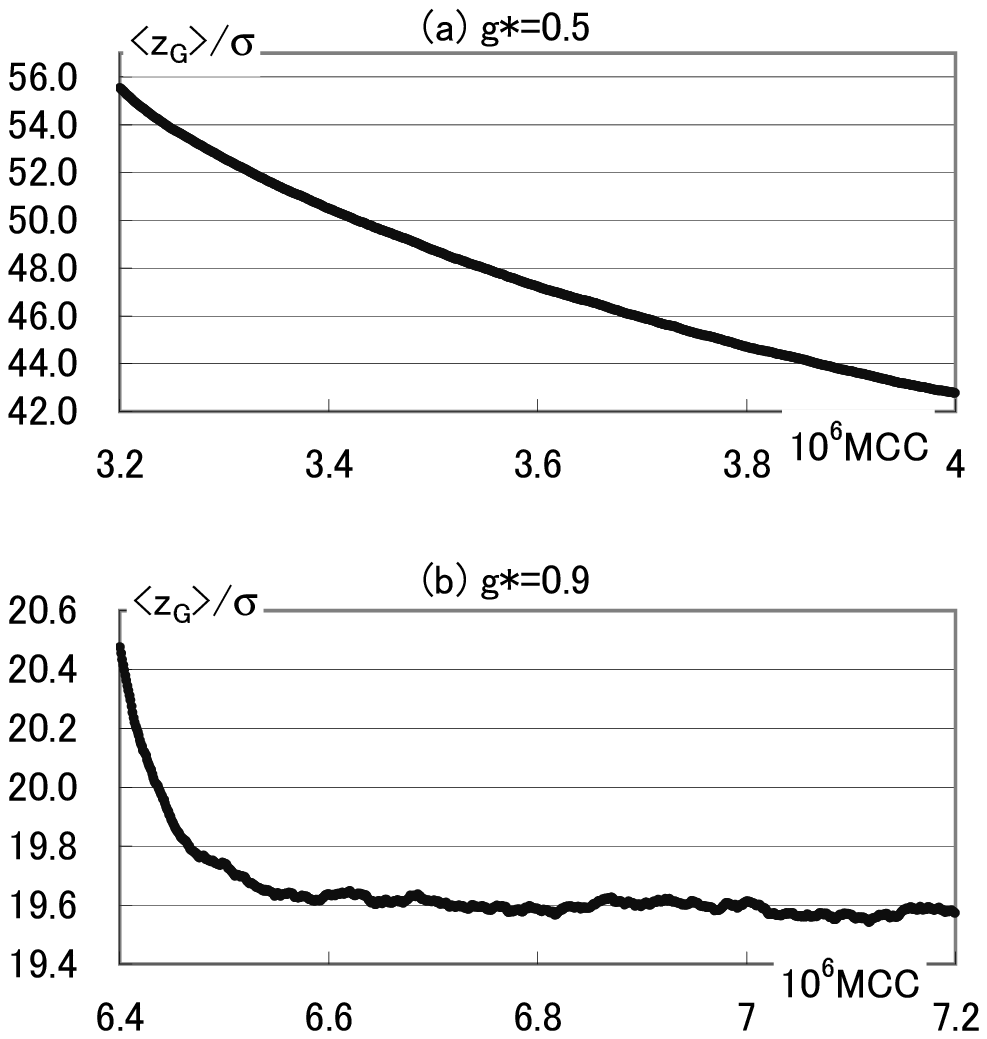}
\end{center}
\caption{Evolution of the center of gravity during (a) $g^*$=0.5 and (b) 0.9 for $N$ = 26624 system.
The both curves in (a) and (b) are of a single relaxation mode.
In (a) relaxation has not reached to the equilibrium yet.
On the other hand, in (b) the lowering of the center of gravity has reached to equilibrium and is fluctuating around it.
Statistical errors are within $0.014\sigma$ for (a) and $0.008\sigma$ for (b).}
\label{fig:gcenter26624}
\end{figure}

Let us look at the evolution of the center of the gravity (Fig.~\ref{fig:gcenter26624}).
Unlike the $N$ = 6656 system, multiple relaxations are not appreciable.
The evolution of the center of gravity during $g^*$ = 0.4 is of a single relaxation mode and does not reached at equilibrium.
Also, that during $g^*$ = 0.9 is of a single relaxation mode.
In this case, however, it has reached at equilibrium and is fluctuating around the equilibrium.
Formation of a defect structure during $g^*$ = 0.9 mentioned above may occur in an equilibrium fluctuation.
However, in a magnification of the equilibrium fluctuation of the evolution of the center of gravity, we can find a correlation of the structural change and the evolution.
Some detail analysis will be given in a future research.

\section{Concluding remarks}
\label{sec:conclu}
We successfully performed Monte Carlo simulations of a colloidal epitaxy on a square pattern using hard spheres.
In other words, we have succeeded in replacing the artificial stress, which is a driving force for fcc (001) stacking, with realizable one.
Moreover, we would say that the system size can be enlarged systematically.
For a large system, however, a number of defects running along deferent directions occurred in a system.
It makes analyses complicated.
 
In a case that defect disappearance was observed at lower $g^*$ than that for the flat bottom wall cases, the sinking of the center of gravity of the system was smooth and of a single relaxation mode.
Also for a large system, it was smooth and of a single relaxation mode.
That is, in this case the shrinking of the defect was not trapped temporarily at a metastable configuration.
On the other hand, at $g^*$ greater than the value at which the defect disappearance and temporal stopping of the lower end of an intrinsic stacking fault occurred for the flat wall cases (at $g^*$ = 0.9), the temporal stopping of the sinking of the center of gravity was observed.
For large system, such temporal stopping was not appreciable.

In the snapshots tetrahedral structures appeared often, suggesting staking fault tetrahedra being sessile.
Observation of the tetrahedral configuration for more large systems are in progress.

To accomplish complicated analyses to observe the manner of defect disappearance and identify the structure of defects are left as future researches.
System size is to be systematically enlarged.
The way of controlling $\Delta g^*$ should be optimized both to observe the details of the defect disappearance and to efficiently erase the defects in reality.





\bibliographystyle{elsarticle-num}
\bibliography{<your-bib-database>}



\end{document}